\documentclass[%
 reprint,
 amsmath,amssymb,
 aps,
 prx,
]{revtex4-2}

\usepackage{graphicx}
\usepackage{dcolumn}
\usepackage{bm}
\usepackage{xcolor}
\usepackage{lineno}

\usepackage{amsmath}
\usepackage{multirow}
\usepackage{soul}
\usepackage{hyperref}

\begin{document}

\preprint{APS/123-QED}

\title{Curvature dependent dynamics of a bacterium confined in a giant unilamellar vesicle}

\author{Olivia Vincent}
\author{Aparna Sreekumari}%
\author{Manoj Gopalakrishnan}
 \email{manojgopal@iitpkd.ac.in}
\altaffiliation[Also at ]{Department of Physics, Indian Institute of Technology Madras, India.}
\author{Vishwas V Vasisht}%
 \email{vishwas@iitpkd.ac.in}
\author{Bibhu Ranjan Sarangi}
 \email{bibhu@iitpkd.ac.in}
\altaffiliation[Also at ]{Department of Biological Sciences and Engineering, Indian Institute of Technology Palakkad, India.}
\affiliation{%
 Department of Physics, Indian Institute of Technology Palakkad, India.
}%


\date{\today}

\begin{abstract}
	We investigate the positional behavior of a single bacterium confined within a vesicle by measuring the probability of locating the bacterium at a certain distance from the vesicle boundary. We observe that the distribution is bi-exponential in nature. Near the boundary, the distribution exhibits rapid exponential decay, transitioning to a slower exponential decay, and eventually becoming uniform further away from the boundary. The length scales associated with the decay are found to depend on the confinement radius. We interpret these observations using molecular simulations and analytical calculations based on the Fokker-Planck equation for an Active Brownian Particle model. Our findings reveal that the small length scale is strongly influenced by the translational diffusion coefficient, while the larger length scale is governed by rotational diffusivity and self-propulsion. These results are explained in terms of two dimensionless parameters that explicitly include the confinement radius. The scaling behavior predicted analytically for the observed length scales is confirmed through simulations.
\end{abstract}

\maketitle


\section{\label{secIntro}Introduction}
Active systems, which are inherently non-equilibrium in nature, consist of components that draw energy from their surroundings to execute systematic movement \cite{ramaswamy2010mechanics}, including directed motion. These components span a wide range of length scales, from molecular motors and bacteria to fish, birds, and other living organisms \cite{ramaswamy2010mechanics, ramaswamy2017active}. The study of active assemblies has been instrumental in advancing our understanding of self-organization and motility-induced phase transitions in the last two decades \cite{elgeti2015physics, gompper20202020, araujo2023steering}. Exploring active matter dynamics in diverse environments has drawn significant interest in recent times. The interactions of active particles with physical obstructions and interfaces are important to understand their collective behavior, sorting, as well as to gain insights about the environment itself \cite{spagnolie2015geometric, bechinger2016active, Sujit_RSC_2023}. Given the complexities of the interactions of active particles among themselves as well as the environment, analyzing the behavior of an individual active particle provides an opportunity to understand active matter systems within the framework of statistical physics \cite{cates2012diffusive, burkholder2019fluctuation, PhysRevE.103.032607}. 

\begin{figure*}[ht]
    \centering
    \includegraphics[width=1.0\textwidth]{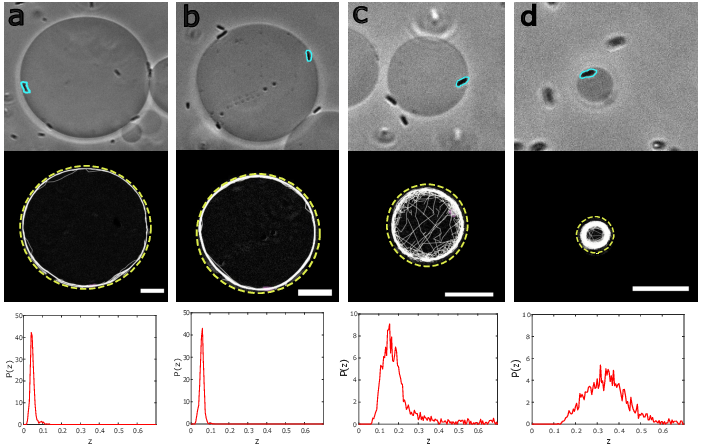}   
     \caption{{\bf Experiments} (top panel) Phase contrast microscopy image of a bacterium (highlighted by the cyan border) encapsulated inside a vesicle of different radius (a) 29$\mu$m, b) 18$\mu$m, c) 8$\mu$m, d) 3$\mu$m). (middle panel) Trajectory of the encapsulated bacterium for the corresponding vesicle radius. The yellow dashed line shows the boundary of the vesicle. The scale bar is 10$\mu$m. (bottom panel) Positional distribution of the bacterium for the corresponding radius. The scaled distance from the circular boundary, z, is defined as $1- r/R$, such that z = 0 is on the boundary and z=1 is at the center.}
    \label{fig1}
\end{figure*}

Motivated by the fact that many active living particles are naturally found in geometrically confined environments, such as blood vessels, fallopian tubes, and porous media such as soil, researchers have investigated their behavior near surfaces, within channels, and in other confined spaces \cite{bechinger2016active, gompper20202020, Sujit_RSC_2023}. Some of the earliest experimental studies aimed at understanding the motion of bacteria, one of the most commonly studied natural active particles, at a planar interface without any flow field, revealing a strong affinity for the surface \cite{frymier1995three, li2008amplified}. The surface affinity has generally been attributed to hydrodynamic interactions \cite{lauga2006swimming, berke2008hydrodynamic, tuson2013bacteria}, however, recent experimental evidence \cite{drescher2011fluid} suggests that in Escherichia coli \textit{(E. coli)} bacteria, hydrodynamic interactions play a relatively minor role. Simulation studies have explored the effect of shape on the surface accumulation of self-propelled particles confined between planar walls \cite{elgeti2009self, elgeti2013wall, elgeti2015physics}. These studies have concluded that even spherical active particles, without hydrodynamics, show surface accumulation that is primarily influenced by the channel width, self-propulsion velocity, translational and rotational diffusion coefficients. In addition to simulations, analytical solutions based on the Fokker-Planck equations have demonstrated that, in the limit of a narrow channel, the positional probability density strongly depends on the P\'eclet number,  which is defined as $P_e = \frac{v_0}{\sqrt{D_r D_t}}$, where $v_0, D_r$ and $D_t$ are the propulsion speed, rotational and translational diffusion coefficients respectively. At low P\'eclet numbers, the density decays exponentially as a function of the distance from the wall, while at higher P\'eclet numbers, it transitions to a power-law decay \cite{elgeti2013wall}. Several experimental studies have provided evidence that, under strong confinement, active Brownian particles (ABP) in a harmonic potential cluster together in a thin ring, at a fixed distance from the center of the potential, corresponding to a strongly non-Boltzmann stationary state \cite{takatori2016acoustic, dauchot2019dynamics, schmidt2021non, buttinoni2022active}. Theoretical studies on curved confinement have demonstrated that the confinement strength can be characterized by the ratio of the radius of curvature to the persistence length of the ABP. Under conditions of zero translational diffusion and relatively strong confinement conditions, the number density of active particles on the curved surface, as well as the associated pressure, is shown to depend on the curvature \cite{fily2014dynamics, fily2015dynamics}. In addition, several other theoretical studies \cite{ hennes2014self, elgeti2015run,  malakar2018steady, duzgun2018active, santra2019universality, basu2019long, kumar2020active, chaudhuri2021active, wagner2022steady,  nakul2023stationary, knippenberg2024motility} have contributed to the current knowledge of active particle within confinement. The affinity of bacteria towards interfaces have  been shown in  experimental studies on confined bacteria inside water droplet \cite{vladescu2014filling}. Using confined bacterium inside a Giant Unilamellar Vesicle (GUV) activity induced perturbations such as shape fluctuations  have also been demonstrated \cite{vutukuri2020active,takatori2020active}. For the case of a softer membrane tether formation by the bacteria was observed which also induced self propulsion in the GUV \cite{le2022encapsulated}. However, a comprehensive understanding of accumulation within confined circular boundaries remains incomplete, particularly in terms of positional density, orientational ordering and their dependence on curvature as well as thermal and active noise. 

In this work, we try to bridge this gap by presenting an in-depth analysis of an active particle within a circular confinement, through a combined approach consisting of experiments, simulations, and analytical calculations. We study a single \textit{ E. coli} bacterium confined within a  (GUV) and measure its positional distribution. Interestingly, the distribution follows a bi-exponential form associated with two distinct length scales. This finding is corroborated by molecular simulations of the Active Brownian Particle (ABP) model, which reveals that the short length scale is influenced by the translational diffusivity whereas the long length scale is dominated by the rotational diffusivity and self-propulsion. Utilising analytical calculations based on the Fokker-Planck equations (FPE), we rationalize these findings in terms of two dimensionless parameters that demonstrate data collapse, indicating a scaling behavior.

The paper is divided into five sections, beginning with the Introduction. The second section details our experimental setup and findings. In the third section, we present the results from simulations of ABP. In the fourth section, we discuss the Fokker-Planck equation and its solution, along with the various approximation procedures involved. This is followed by a discussion of our results in the final section. Experimental and mathematical details are provided in two appendices. 

\section{\label{secExp} Bacterium inside a Vesicle}
To study the motion of \textit{ E. coli} bacterium under confinement, we encapsulate a living bacterium within a Giant unilamellar vesicle (GUV) by utilising a modified electroformation technique \cite{angelova1986liposome, kuribayashi2006electroformation, takatori2020active}. To form the GUVs we use 1,2-Dioleoyl-sn-glycero-3-phosphocholine (DOPC) and typically DOPC-composite lipid bilayer have a rigidity of around  $~20 k_BT$ \cite{zabala2023determining, karal2023review}. We have produced GUVs with varying radius (R) values ranging from $3\mu$m to $30\mu$m. Our experimental protocol is optimized to precisely control the number density of bacteria inside the GUVs and in all the experimental data reported in this work involves a single bacterium confined within a GUV (see Appendix A).

For each GUV, we obtain the positional coordinates of the confined bacterium using phase-contrast microscopy. The time-lapse images are analyzed using an open source image tracking algorithm - ImageJ \cite{ershov2022trackmate}. We observe that the bacterium has an affinity towards the inner surface of the vesicle and moves around the concave boundary, in clockwise or anticlockwise manner with occasional flipping around. This is expected behavior for an active particle confined within an enclosure. A dead bacterium does not show such a behavior and its motion is similar to a passive Brownian particle (see Movie 2 and Fig.1, SI). In our experiments, the majority of bacterial trajectories were confined to the equatorial plane of the GUV. Notably, the GUVs exhibited a slight deviation from perfect sphericity, adopting a mildly oblate geometry. This subtle asymmetry may bias the bacteria to preferentially explore the equatorial plane thereby constraining its trajectory to a planner one. Accordingly, all our experimental data pertain to these trajectories, which can be considered effectively two-dimensional. In Fig.~\ref{fig1} top panel (also Movie 1, SI) we show the phase contrast microscopy image of a bacterium enclosed within a vesicle for four different radius of curvature. The trajectories of the bacterium is shown in the middle panel, which shows that the affinity toward the boundary decreases with a decrease in $R$. We compute the positional distribution $P(z)$, shown in the bottom panel. Here z is the scaled distance from the circular boundary, defined as $z = 1 - (r/R)$ with $r$ being the radial distance of the bacterium from the center of the vesicle, such that z = 0 is on the boundary and z=1 is at the center.
 $P(z)$ initially increases with $z$, near the boundary, reaches a peak, and then decreases as $z$ continues to increase.  The position of the peak for all bacteria is observed with a constant offset from the boundary, which can be attributed to the finite size of the bacterium. Next, we provide a detailed analysis of $P(z)$.

\begin{figure}[th]
    \centering
    \includegraphics[width=1.0\columnwidth]{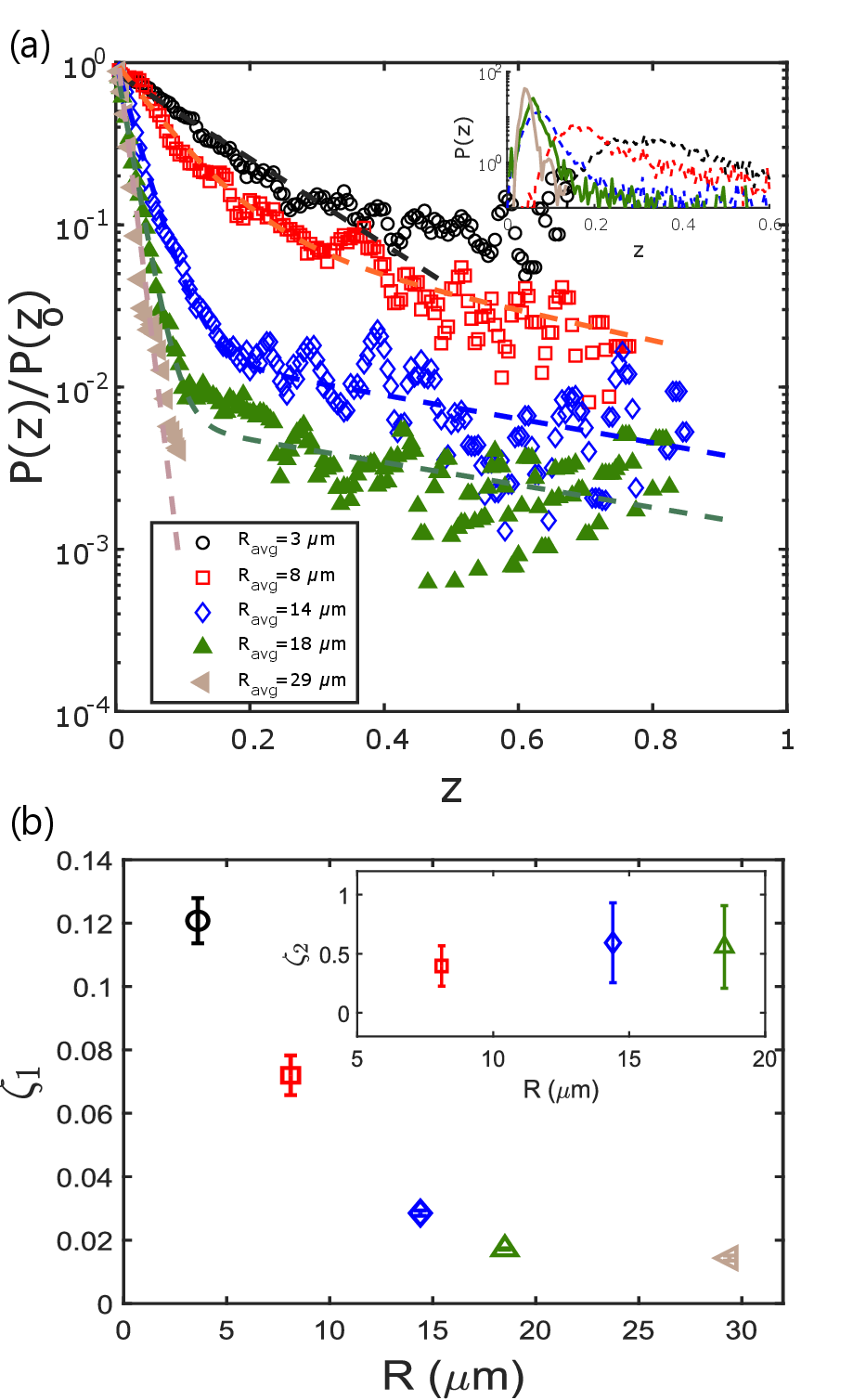}   
    \caption{{\bf Experiments} (a) (main panel) Scaled positional distribution function for different vesicle radius. The dashed lines correspond to bi-exponential fits for selected data. (inset) Full positional distribution. (b) The length scales $\zeta_1$ (main panel) and $\zeta_2$ (inset) extracted from a bi-exponential fitting function. Error bars indicate 95\% confidence intervals, calculated from the upper and lower confidence bounds of the fit}.
	\label{fig2}
\end{figure}

\noindent {\bf Positional distribution : } In Fig.~\ref{fig2} (a) main panel, we show the scaled positional distribution (beyond the peak value) for different $R$ values. Here, $P(z_0)$ denotes the peak value of the distribution. The distribution is obtained by averaging multiple independent experimental measurements, with $R_{avg}$ representing the average $R$ binned over a range of vesicle radii (see Fig.~\ref{fig2} (a) inset for a full distribution). The experimental data fits quite well to the bi-exponential function $a_0 e^{-z/\zeta_2}+(1-a_0)e^{-z/\zeta_1}$ (shown as dashed line in Fig.~\ref{fig2} (a)). As $R$ decreases, we observe a slower decay. The associated length scales $\zeta_1$, and $\zeta_2 $ are extracted and plotted as functions of $R$ in Fig.~\ref{fig2} (b). The short length scale $\zeta_1$ monotonically decreases with an increase $R$ before saturating. The long length scale $\zeta_2$, which is clearly evident for $R_{avg}=18 \mu$m and $R_{avg}=14 \mu$m, remains fairly constant with increasing $R$ (Fig.~\ref{fig2} (b) inset).

 To better understand the experimental observations, we perform molecular simulations in which the bacterium is modeled as an ABP.  The circular confinement is introduced as a reflecting boundary condition on the positional coordinates. For simplicity, we do not include hydrodynamics explicitly, as hydrodynamic effects are considered to have a weak influence on the affinity of ABP to the wall \cite{vladescu2014filling, elgeti2009self, drescher2011fluid, elgeti2013wall}. 
 
\section{\label{secSim} Active Brownian Particle Model}

Following the simulation protocol as outlined in \cite{mandal2020extreme}, we numerically solve the Langevin equations of motion  \cite{allen2017computer}.
\begin{eqnarray}
m{\ddot{{\bf r}}}&=&-\gamma_t({\dot{{\bf{r}}}}-u_0{\hat {\bf u}})+{\boldsymbol{\nu}_t},\label{eq1}\\
{\dot{\theta}}&=&{\nu_r}\label{eq2}, 
\end{eqnarray}
where $\gamma_t$ is the viscous drag coefficient for translational motion, ${\boldsymbol{\nu}}_t(t)$ is the translational thermal noise term with zero mean and autocorrelator 
$\langle {\boldsymbol{\nu}}_t(t)\cdot{\boldsymbol{\nu}}_t(t^{\prime})\rangle=4{k}_{B}T\gamma_t\delta (t-t^{\prime})$. The thermal translational diffusion coefficient is $D_t=k_B T/\gamma_t$ from the fluctuation-dissipation theorem, where $T$ is the absolute temperature. Here, $u_0{\hat {\bf u}}$ is the active velocity, whose direction is specified by the unit vector ${\hat{\bf u}} = (\cos \theta, \sin\theta)$. The Gaussian white noise $\nu_r(t)$ that controls rotational motion also has zero mean and its autocorrelator is given by $\langle \nu_r(t)\nu_r(t^{\prime}) \rangle = 2D_r\delta(t-t')$, where $D_r$ is the rotational diffusion coefficient. The persistence time of the ABP is $\tau_P=D_r^{-1}$ and the persistence length $l_P=u_0/D_r$. In Eq.~\ref{eq1}, we set $\gamma_t=10$ which drives the particle's positional evolution into an overdamped regime while the angular evolution is overdamped as per Eq.~\ref{eq2}. Although exploring inertial effects on confined particles is an interesting topic \cite{caprini2022role, sprenger2023dynamics}, it falls outside the scope of this study. For the simulation results presented in this work, the parameters used (all in reduced units) are as follows: $m=1$, $u_0=0.25$, and the integration time step $\Delta t=0.01$. We set $T=0.001$ and hence $D_t = 10^{-4}$ (unless explicitly stated otherwise), and $D_r$ varies between $10^{-3}$ and $10^{-1}$. The confinement radius $R$ varies between 1 and 50. For the purpose of direct comparison with the experiments and the theoretical predictions, certain dimensionless ratios of the above parameters will be introduced later. The Brownian dynamics code \cite{allen2017computer} was modified to incorporate activity noise for the simulation data presented in this work. Sufficient statistics were obtained by performing approximately 10 million $\tau_P$ steps and collecting 10 independent samples at each state point.

\begin{figure}[th]
    \centering  \includegraphics[width=1.0\columnwidth]{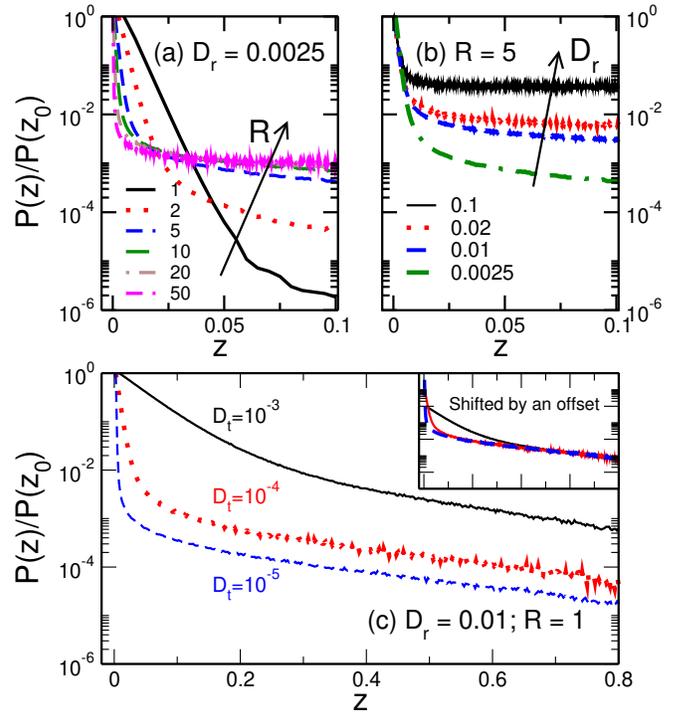}   
	\caption{{\bf Simulations} (a) Scaled positional distribution for fixed $D_r=0.0025$ and varying confinement radius ($R=$1, 2, 5, 10, 20 and 50). (b) Scaled positional distribution for fixed $R=5$ and varying rotational diffusion ($D_r=$0.1, 0.02, 0.01 and 0.0025). (c) Scaled positional distribution for fixed $D_r=0.01$ and $R=1$ and varying $D_t$ ($10^{-3}, 10^{-4}$ and $10^{-5}$). To highlight that the distribution becomes independent of $D_t$ at larger $z$ values, the curves have been shifted relative to one another, using one curve as a reference, and shown in the inset}
    \label{fig3}
\end{figure}

\noindent{\bf Positional distribution:} In Fig.~\ref{fig3} (a) we show the positional distribution $P(z)$, computed from numerical simulations for a fixed $D_r$ and varying $R$ in a log-linear plot. The distribution is scaled by $P(z_0)$, where $z_0$ represents either the peak value of the distribution or the distribution value at the initial point in $z$. The distribution shows not only the primary exponential decay but also the slow secondary decay, which are consistent with our experimental observations. As we move further away from the boundary, towards the center, $P(z)$ slowly tends to a nearly uniform distribution. The $R$-dependence of $P(z)$ aligns with our experimental findings, showing a slower decay with a decrease in $R$. While manipulating bacterial activity is challenging in experiments, it is easily achievable in simulations by tuning the persistence time or $D_r$. Therefore, we decided to investigate the effects of $D_r$ alongside $R$. In Fig.~\ref{fig3} (b) we show $P(z)$ for varying $D_r$ at a fixed $R$. For a wide range of $D_r$ values, the observed double exponential nature of the decay remains consistent. As $D_r$ increases, indicating an effective reduction in activity, the second exponential decay region diminishes, and the distribution becomes uniform rapidly. We also investigated the positional distribution $P(z)$ for two other values of $D_t$ ($10^{-3}$ and $10^{-5}$), as shown in Fig.~\ref{fig3} (c). We observe that, as $D_t$ decreases, the primary exponential decay is quicker. In the inset of Fig.~\ref{fig3} (c), we present $P(z)$ with an offset, using a reference point at larger $z$. This demonstrates that the secondary exponential decay remains unaffected by the change in $D_t$.

\begin{figure}[th]
    \centering  \includegraphics[width=1.0\columnwidth]{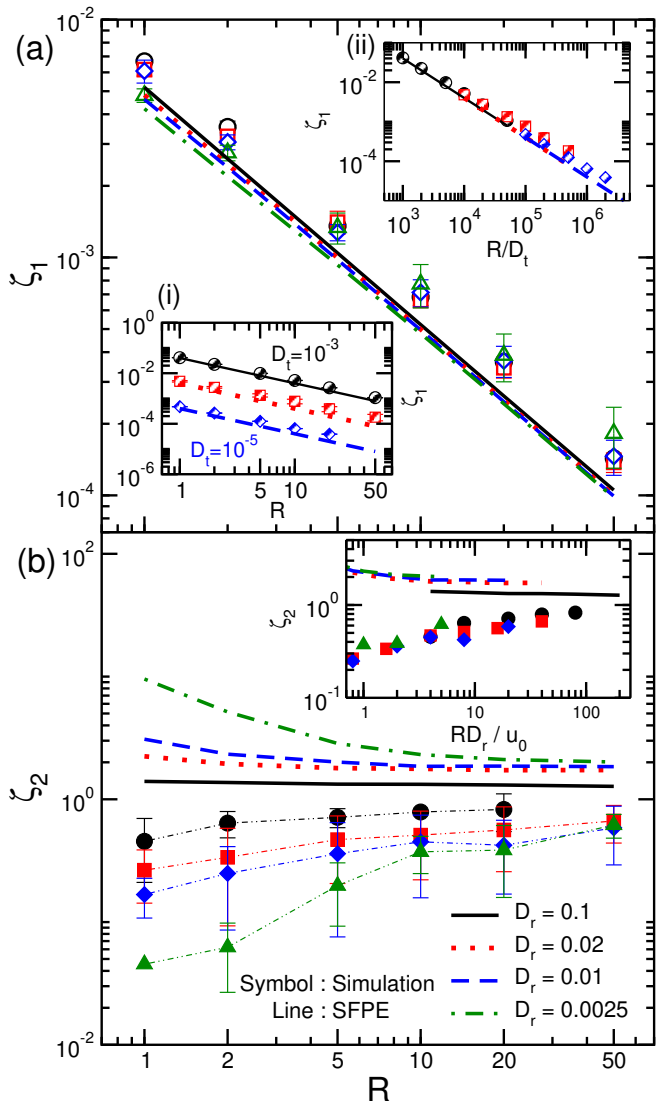}   
	\caption{{\bf Length scales from simulation (symbols) and SFPE (lines)} (a) The small length scale $\zeta_1(R)$ for fixed $D_t=10^{-4}$ and varying $D_r=0.1, 0.02, 0.01, 0.0025$. Inset (i): For a fixed $D_r=0.0025$, $\zeta_1(R)$ is shown for $D_t=10^{-3}$ (filled circle), $10^{-4}$ (filled square) and $10^{-5}$ (filled diamond). In the inset (ii) we show the same data, with $R$ scaled by $D_t$. (b) Simulation data for the larger length scale $\zeta_2$ is shown, in comparison with the theoretical predictions. In the inset we show the same data, with $R$ scaled by $u_0/D_r$, which is the persistence length $l_P$ of the ABP (see text for explanation).}
    \label{fig4}
\end{figure}

\noindent{\bf Length scale dependence on $R$, $D_r$ and $D_t$ : } The above observations are effectively captured by extracting the two length scales, $\zeta_1$ and $\zeta_2$. We have extracted the length scales both from bi-exponential fit as well as from piece-wise exponential fits to the data (which provides an error bar). The symbols in the main panel Fig.~\ref{fig4} (a) shows the variation of $\zeta_1$ with $R$ for fixed $D_t=10^{-4}$ and varying $D_r$. $\zeta_1$ decreases with increasing $R$, following a power law, with an exponent $\approx -1$ and shows a little or no dependence on $D_r$. In Fig.~\ref{fig4} (a) inset (i) we show $\zeta_1(R)$ for fixed $D_r=0.0025$ and varying $D_t$, highlighting a clear dependence on $D_t$. In contrast, $\zeta_2$ increases with $R$ before reaching a saturation point and shows a clear dependence on $D_r$ (see Fig.~\ref{fig4} (b) symbols). In the inset of Fig. ~\ref{fig4} (b) we show that $\zeta_2$, for varying $D_r$, collapse in a single profile as we scale $R$ by the persistence length $l_P$.

{The above simulation results demonstrate that a simple ABP model exhibits the bi-exponential positional distribution observed in our experiments. The extracted length scales are distinctly dependent on the translational and rotational diffusion coefficients. In addition,  $\zeta_1$ is found to be inversely proportional to $R$ and $\zeta_2$ increases with increase in $R$, before saturating. We now proceed to develop the Fokker-Planck equation based analytical formalism to gain deeper insights into the observed positional distribution and the associated length scales.  

\section{\label{secTheo} Theory: Fokker-Planck equation}

In the over-damped regime where inertial effects are unimportant, the dynamics of the position variable ${\bf r}$ is described by a simpler first-order equation, which follows from Eq.~\ref{eq1}: 
\begin{equation}\label{eq3}
    {\dot{{\bf{r}}}} = u_{0} \, \hat{\mathbf{u}}(t) + \sqrt{2D_t}{\boldsymbol {\nu}}_t \,
\end{equation}
Eq.~\ref{eq2} and Eq.~\ref{eq3} constitute the complete over-damped dynamical equations for the ABP. The interaction of the particle with the boundary is incorporated in the Fokker-Planck equation through a reflecting boundary condition for the positional distribution. 

We represent by $(z,\phi)$, the position of the particle in plane polar coordinates. On account of the rotational symmetry of the confinement geometry, the stationary state distribution does not explicitly depend on $\phi$, and the dependence on $\theta$ occurs through the combination $\chi=\theta-\phi$ \cite{malakar2018steady}. Let $f(z,\chi)$ be the joint probability density for the position and orientation of the particle.  When the activity is sufficiently large, it is expected that the particles will mostly stay close to the boundary, in which case we may assume that  $z\ll 1$. In this limit, we show that the probability density satisfies the following linearized stationary FPE (SFPE): 
\begin{equation}
(1+\eta)\frac{\partial^2 f}{\partial \chi^2}+\frac{\partial j}{\partial z}+{\tilde u}\frac{\partial}{\partial \chi}(\sin\chi f)-j\simeq 0,
\label{eq4}
\end{equation}
where 
\begin{equation}
j(z,\chi)={\tilde u}\cos\chi f+\eta\partial f/\partial z
\label{eq4+}
\end{equation}
is the radial component of the translational probability current density (Eq.~\ref{eq17}, Appendix B), and 
\begin{equation}
{\tilde u}=\frac{u_0}{RD_r}~~~;~~~~\eta=\frac{D_t}{R^2 D_r},
\label{eq5}
\end{equation}
are two useful dimensionless ratios. Here $\tilde u$ is the ratio of persistence length to radius of curvature and hence represents the confinement strength.  $\eta$ is the measure of the relative significance of thermal translational noise in comparison to rotational noise. The experimental values of $\eta$ are found to be in the range $7.8\times 10^{-4} - 5.2\times 10^{-2}$, where the smallest value corresponds to $R=29\mu$m and the largest value, to $R=3.25\mu$m (see Appendix A). In our mathematical analysis, we shall, therefore, generally assume $\eta\ll 1$ for the sake of simplification, wherever it is convenient to do so. 

The marginal distributions for the position coordinate $z$ and the orientation angle $\chi$ are defined as follows: 
\begin{equation}
P(z)=\int_{-\pi}^{\pi}d\chi f(z,\chi)~~;~~~g(\chi)=\int_{0}^{1}f(z,\chi)(1-z)dz
\label{eq6}
\end{equation}

We solve Eq.~\ref{eq4} after incorporating the purely reflecting boundary condition at the wall, i.e., $j(0,\chi)=0$. This is done in a convenient way using Laplace transforms, however since $z\in [0:1]$, quantitatively accuracy in results is expected only when $f(z,\chi)$ decays very fast away from $z=0$ (the boundary). 

\subsubsection{Laplace transform-based solution of Eq.~\ref{eq4}}

For $n=0,1,2...$, let us define a set of functions 
\begin{equation}
\lambda_n(z)= \int_{-\pi}^{\pi}d\chi \cos^n\chi f(z,\chi)\equiv \alpha_n(z)P(z)
\label{eq19}
\end{equation}
where $\alpha_0(z)=1$ and 
\begin{equation}
\alpha_n(z)= \langle \cos^n \chi\rangle_z~~~~~(n\geq 1)
\label{eq19+}
\end{equation}
are conditional moments of $\cos\chi$. We also define a set of Laplace transforms 
\begin{equation}
{\tilde \lambda}_n(s)=\int_{0}^{\infty}\lambda_n(z)e^{-sz}dz~~~~(n=0,1,2,....)
\label{eq20}
\end{equation}
Eq.~\ref{eq4} is now subjected to Laplace-transformation with respect to $z$. After implementing the reflecting boundary condition $j(0,\chi)=0$, we arrive at the equation

\begin{equation}
(s-1){\tilde j}(s)+(1+\eta)\frac{\partial ^2{\tilde f}}{\partial \chi^2}+{\tilde u}\frac{\partial}{\partial \chi}(\sin\chi {\tilde f})=0
\label{eq21}
\end{equation}
where the Laplace-transformed current density is 
\begin{equation}
{\tilde j}(s,\chi)=(\eta s+{\tilde u}\cos\chi){\tilde f}-\eta f(0,\chi).
\label{eq22}
\end{equation}
Next, we substitute Eq.~\ref{eq22} in Eq.~\ref{eq21} and arrive at the complete linearized SFPE in Laplace space: 

\begin{eqnarray}
(1+\eta)\frac{\partial^2{\tilde f}}{\partial \chi^2}+{\tilde u}\frac{\partial}{\partial \chi}(\sin\chi {\tilde f})+\nonumber\\
(s-1)({\tilde u}\cos\chi+\eta s){\tilde f}=\eta(s-1)f(0,\chi).
\label{eq23}
\end{eqnarray}
To arrive at an equation for the positional distribution, we integrate Eq.~\ref{eq23} over $\chi$, and use the definitions of the marginal quantities as given in Eq.~\ref{eq6}. The first two terms on the l.h.s vanish on account of periodicity of ${\tilde f}$ with respect to $\chi$. Next, we multiply Eq.~\ref{eq23} with $\cos\chi$ and integrate over $\chi$. As a result, the transforms defined in Eq.~\ref{eq20} are found to be related through an infinite hierarchy of equations. The first two of these turn out to be  

\begin{eqnarray}
\eta s{\tilde P}(s)=\eta P(0)-{\tilde u}{\tilde \lambda}_1(s),~~~~~~\label{eq24}\\
\left\{\eta s(s-1)-(1+\eta)\right\}{\tilde \lambda}_1(s)=\eta(s-1)\lambda_1(0)-{\tilde u}{\tilde P}(s)-\nonumber\\
{\tilde u}(s-2){\tilde \lambda}_2(s).~~~~~~.
\label{eq25}
\end{eqnarray}
As an important consistency check, note that, as ${\tilde u}\to 0$, Eq.~\ref{eq24} gives ${\tilde P}(s)\to P(0)/s$, which implies $P(z)\to {\rm constant}$. On the one hand, this is only expected, since, when $u_0=0$, the ABP becomes a passsive Brownian particle whose stationary distribution inside a hard-wall confining well is uniform. Our scaling analysis predicts a more general condition $l_p\ll R$, for the active-passive crossover, where $l_p=u_0/D_r$ is the persistence length. This is supported by simulation data (see Fig.~\ref{fig3}, a and b).

\subsubsection{A semi mean-field approximation for the orientation variable}

The first two moments of $\cos\chi$, defined in Eq.~\ref{eq19+} are crucial in the determination of the positional probability $P(z)$. However, the equations determining these moments are not closed. Therefore, we now devise a suitable closure scheme, which, although approximate, allows us to make further analytical progress. The essence of the approximation is to first write $\alpha_2(z)=\alpha_1(z)^{2}+\sigma^2(z)$, where the last term is the conditional variance of $\cos\chi$. We now invoke a partial mean-field approximation for $\alpha_2$, whereby the variance is entirely neglected, and the conditional average is replaced by the corresponding global mean, i.e., 
$\alpha_2(z)\simeq \alpha_1(z)^{2}$ followed by $\alpha_1(z)\to \overline{\alpha}_1$, leading to 
\begin{equation}
\lambda_2(z)\simeq \overline{\alpha}_1^2 P(z)~~~~~~~({\rm semi~mean-field})
\label{eq25++}
\end{equation}
where
\begin{equation}
\overline{\alpha}_1\equiv \int g(\chi)\cos\chi d\chi
\label{eq27}
\end{equation}
is the global average of $\cos\chi$, and the marginal distribution $g(\chi)$ is defined in Eq.~\ref{eq6}. From the above definition, the exact relation 
\begin{equation}
{\tilde \lambda}_1(0)=\overline{\alpha}_1{\tilde P}(0)
\label{eq27+}
\end{equation}
also follows (Eq.~\ref{eq19} and Eq.~\ref{eq20}). We emphasize that the spatial variation in the first moment is retained in our equations, and hence the term semi-mean-field. 

The global mean $\overline{\alpha}_1$ can be determined self-consistently within the semi-mean-field approximation. To do this, we start with Eq.~\ref{eq25}, substitute ${\tilde \lambda}_2(s)={\overline{\alpha}}_1^2 {\tilde P}(s)$  and put $s=0$. After using the relation in Eq.~\ref{eq27+}, we arrive at the equation
\begin{equation}
[2{\tilde u}{\overline{\alpha}}_1^{2}+(1+\eta){\overline{\alpha}}_1-{\tilde u}]{\tilde P}(0)=\eta\lambda_1(0)
\label{eq28}
\end{equation}
It follows that, in the limit $\eta\to 0$, ${\overline{\alpha}}_1$ satisfies the quadratic equation $2{\tilde u}{\overline{\alpha}}_1^2+{\overline \alpha}_1-{\tilde u}=0$. Among the two (real)  solutions, only one, i.e., 
\begin{equation}
{\overline{\alpha}}_1^{\rm SMF}=\frac{-1+\sqrt{1+8{\tilde u}^2}}{4{\tilde u}}~~~~(\eta\to 0)
\label{eq29}
\end{equation}
is physically meaningful, since ${\overline{\alpha}}_1\in [-1:1]$ by definition (the magnitude of the negative root is greater than 1). The expression in Eq.~\ref{eq29} is a monotonically increasing function of ${\tilde u}$. As ${\tilde u}\to 0$, ${\overline{\alpha}}_1^{\rm SMF}\simeq {\tilde u}$ while ${\overline{\alpha}}_1^{\rm SMF}\to 1/\sqrt{2}$ as ${\tilde u}\to \infty$. The positive value of ${\overline{\alpha}}_1^{\rm SMF}$ suggests the existence of strong orientation order for large ${\tilde u}$, and we verified this using numerical simulation data. The predicted saturation value in the limit of large ${\tilde u}$ is less than 1, which contradicts simulation data. We believe that this discrepancy indicates a limitation of the mean-field nature of the theory.

\begin{figure}[htp!]
	\centering  \includegraphics[width=1\columnwidth]{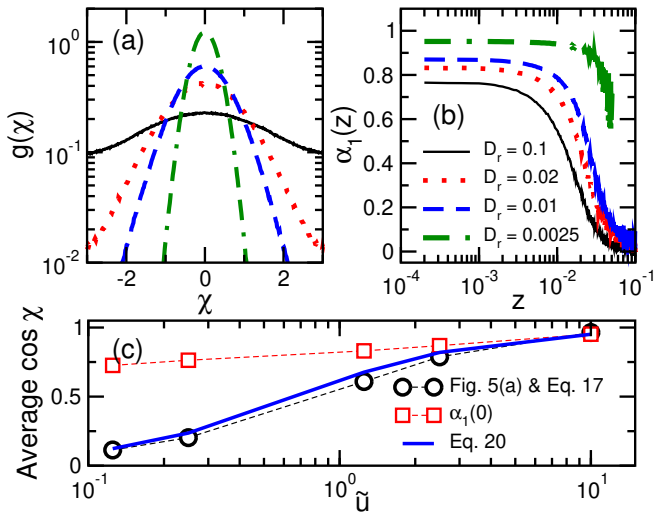}   
	\caption{ (a) Angular distribution $g(\chi)$ (Eq.~\ref{eq6}) for varying $D_r$ at fixed $R=1$.  (b) The conditional average of $\cos \chi$ (Eq.~\ref{eq19+}) as a function of $z$ for varying $D_r$ and fixed $R=1$. Both (a) and (b) are obtained from simulations. (c) Average of $\cos \chi$, from three different approaches, i.e.,  (i) the global mean computed using $g(\chi)$ in (a) (open circles, black) (ii) the $z\to 0$ limit of the conditional mean in (b) (open squares, red) (iii) the semi-mean-field expression for the  global mean in Eq.~\ref{eq29}, for varying $\eta$. Here the data is shown only for $R=1$, with $D_r$ varying.}
	\label{figsimtheory}
\end{figure}

Fig.~\ref{figsimtheory}(a) shows (for $R=1$) the numerically obtained distribution $g(\chi)$ for a few different values of $D_r$, while Fig.~\ref{figsimtheory}(b) shows the corresponding results for the conditional mean $\alpha_1(z)$ (Eq.~\ref{eq19+}). Note that the conditional mean has a reverse-sigmoidal form and shows a high degree of orientational order ($\alpha_1(z)\simeq  1$) close to the boundary. For a considerable range of $z$, $\alpha_1(z)$ remains constant, positive and close to unity, thus driving the motion of the active particles towards the wall, before it abruptly decays to zero. In addition, the plateau part of $\alpha_1$ reduces with increasing $D_r$, effectively signaling a reduction in activity. Therefore, the observed orientational order is closely related to the activity of the ABP.

The similarity between the conditional average $\alpha_1(z)$ and the global average $\cos\chi$ is also assessed from the simulation data. Fig.~\ref{figsimtheory}(c) shows a direct comparison between the two, where the boundary value $\alpha_1(0)$, representing the plateau part in Fig.~\ref{figsimtheory}(b), and the corresponding global mean $\overline{\alpha_1}$ (computed from Eq.~\ref{eq27}, using the numerically obtained distribution in Fig.~\ref{figsimtheory}(a) are plotted together. It is observed that $\alpha_1(0)$ compares favourably with the global mean at larger $\eta$ values, but deviates at lower $\eta$. The semi-mean-field prediction for the global average (Eq.~\ref{eq29}) is also shown for comparison, and is found to agree quite well with the corresponding simulation data. It is worth noting that the mean-field 
prediction (as well as the simulation result) for the global mean under-estimates the local orientational order close to the boundary, but shows a similar behavioral trend as functions of $\eta$. Unless otherwise specified, for purposes of comparison with simulation and experimental data, we have consistently used $\alpha_1(0)$ in place of the global mean, as the former gives better results.

\subsubsection{Solution for $P(z)$}
Having thus established the utility of the linearized SFPE and the semi-mean-field approach-based simplification of the same, we now proceed further and derive the explicit mathematical form for $P(z)$. After substituting Eq.~\ref{eq25} in Eq.~\ref{eq24}, combining the two, and using the semi mean-field approximation in Eq.~\ref{eq25++}, we arrive at 
\begin{equation}
{\tilde P}(s)=\eta\frac{\left[A(s)P(0)-{\tilde u}s\lambda_1(0)\right]}{\Gamma(s)} 
    \label{eq7}
\end{equation}
where 
\begin{equation}
\Gamma(s)=\eta^2(s^3-s^2)-({\tilde u}^2{\overline\alpha}_1^2+\eta+\eta^2)s-[{\tilde u}^2(1-2{\overline\alpha}_1^2)]
\label{eq30}
\end{equation}
is a cubic polynomial in $s$, whose zeroes determine the behaviour of $P(z)$, $A(s)=\eta(s^2-s)-(1+\eta)$. ${\overline \alpha}_1$ is given by Eq.~\ref{eq29} within the mean-field approximation, or may be borrowed directly from the simulation data as $\alpha_1(0)$. 

A graphical analysis, combined with intuitive arguments shows that the three roots of the cubic equation $\Gamma(s)=0$ are all real and may be denoted $s_0, s_1,s_2$, with $s_0,s_1<0$ and $s_2>0$. Under some conditions, the inverse Laplace transform can be expressed in terms of the two negative roots, and has the bi-exponential form. 

\begin{equation}
P(z)\simeq C e^{-z/\zeta_2}+De^{-z/\zeta_1}, 
\label{eq9}
\end{equation}
where $\zeta_2=1/|s_0|$ and $\zeta_1=1/|s_1|$, and $C$ and $D$ are two numerical pre-factors. For the sake of comparison between theoretical predictions and experimental/simulation results, the roots $s_0$ and $s_1$ are computed numerically using a Newton-Raphson method.

In Fig.~\ref{fig4}, we have shown direct comparisons between the length scales obtained from the simulation data and the corresponding theoretical predictions for the same. For the smaller length scale, we observe the robust scaling behaviour $\zeta_1\propto R^{-1}$ in both simulations and theory, and the pre-factors too agree reasonably well with each other. This suggests that, in the original units, the corresponding length scale $R\zeta_1$ is independent of the radius of curvature and may be considered as an intrinsic, microscopic length scale that characterizes the clustering of non-interacting ABP at the boundary. The length scale $\zeta_2$, on the other hand, approaches a constant value as $R\to\infty$, which suggests that, in the original units, the corresponding length scale increases with the system size. As far as functional dependence on $R$ is concerned, simulation data and theoretical prediction show similar trends. The larger discrepancy between theory and simulations at small $R$ may be attributed to the higher error associated with linearization of the SFPE in this limit.

The difference in the nature of $\zeta_1$ and $\zeta_2$ may be explained in a semi-quantitative way as follows. In Eq.~\ref{eq33}, Appendix B, we derive the following analytical estimates for the roots: $s_0\sim 1-1/{\overline\alpha}_1^{2}$ and $s_1\sim -({\tilde u}/\eta){\overline\alpha}_1$, in the limit $\eta\to 0$. From Eq.~\ref{eq5}, it follows that $\zeta_1\sim (u_0{\overline \alpha}_1/D_t) R$, which also follows naturally from the reflecting boundary condition at the wall (see Appendix B). If the mean-field average ${\overline \alpha}_1$ is replaced by the boundary value $\alpha_1(0)$  (obtained from simulations) for better accuracy, which shows only weak dependence on $R$ (Fig.~\ref{figsimtheory}(c)), we find $\zeta_1\propto (R/D_t)^{-1}$. Experimental data are  consistent with the $1/R$ dependence of $\zeta_1$ (see Fig.S2, SM). Fig.~\ref{fig4} (a) Inset (ii) demonstrates the scaling behaviour of $\zeta_1$ convincingly. The argument also shows that $\zeta_2$, being dependent only on ${\tilde u}$ through ${\overline{\alpha}}_1$, should not have a strong $R$-dependence. This is corroborated by simulation data (Fig.~\ref{fig4}(b)). The inset of Fig.~\ref{fig4} b shows $\zeta_2$ plotted against ${\tilde u}^{-1}=R/l_p$, and we observe the expected collapse of the data, thus confirming the scaling behavior predicted by the theory.

In Table 1, SM, we summarize the experimental data for the two length scales (Fig.~\ref{fig2}(b)) and added comparisons with the corresponding linearized SFPE-based theoretical predictions from the ABP model. For $\zeta_1$, the theoretical predictions under-estimate the observed values by almost an order of magnitude. Since we have found close agreement between the former and simulation data, it appears likely that this discrepancy indicates the limitations of the ABP model in describing bacterial motion. It is possible that hydrodynamic effects and/or tumbling of bacteria \cite{solon2015active, elgeti2015run} significantly affect the positional distribution close to the boundary wall. Indeed, we find better quantitative agreement between theory and experiments for $\zeta_2$, which describes the decay of $P(z)$ away from the boundary. In general, the agreement between theoretical predictions and experimental values is better at higher $R$ values, as expected. For $R\simeq 18.5\mu$m, for example, the experimental value for $\zeta_1$ is 0.019, while the best theoretical estimate is 0.001, smaller by a factor of 3. For $\zeta_2$, the corresponding numbers are 0.558 and 1.20 respectively, and here, our theory over-estimates the observed value by a similar factor.

\section{Discussion and Conclusions}
In this work, we have explored the behavior of a naturally occurring active particle, a bacterium, within a GUV, using a combination of experiments, simulations, and analytical calculations. The positional distribution of the bacterium inside a GUV demonstrates its strong affinity towards the inner boundary of the vesicle. We have studied the nature of the distribution, and find that the probability of locating the bacterium at a specific distance from the boundary is best characterized by a bi-exponential function. This observation is reproduced in our simulations based on a simple minimal model, i.e., an ABP trapped within a circular confinement with a reflective boundary. Mathematical results obtained from a linearized Fokker-Planck equation also predict the nature of the distribution within the constraints of the simplifying assumptions. 

The simulation results reveal some interesting facts about the nature of the two length scales that characterize the bi-exponential decay of the positional distribution. The short-length scale appears to be directly linked to the translational diffusion associated with thermal noise. In contrast, the long-length scale is controlled solely by the rotational diffusion associated with the active noise. Our mathematical formalism predicts specific scaling behaviours for both length scales, which are confirmed in simulations.

 The results of numerical simulation, as well as our mathematical theory, show that the nature of decay of the positional distribution is closely associated with the spatial dependence of the statistics of the orientation variable. The mean orientation of the particles is predominantly radially outward, close to the boundary, and remains constant over a certain region near the wall. Away from the boundary layer, the mean orientation abruptly decays to zero, and the positional distribution switches to the slower mode of decay. This suggests that, unlike what might be expected naively, the ABP orientation does not become disordered beyond the regime of fast exponential decay. The orientational profile of the ABP as a function of distance, is part of our ongoing investigations.

Among existing studies,\cite{duzgun2018active} has investigated  the spatial distribution and orientation of ABPs and their circular hard-wall confinement. The authors derived an analytic solution for the stationary particle density and the orientational order parameter as a function of distance from the center. However, the FPE has a simpler form than our Eq.~\ref{eq14}, which can possibly be attributed to an incomplete treatment of  the orientation-position coupling. The particle density as a function of the radial coordinate was predicted to be given by the modified Bessel function $I_0(r/\ell)$ where $\ell \sim \sqrt{Dt/D_r}$ becomes the diffusive length scale \cite{elgeti2015run} in the limit of low activity ($u_0\ll \sqrt{D_r D_t}$), and $\ell\sim D_t/u_0$ in the opposite limit of high activity. In dimensionless units, the corresponding prediction $\ell/R$ matches our prediction for $\zeta_1$ when $\alpha\simeq 1$, which in our framework, is realized when ${\tilde u}\gg 1$, or $u_0\gg RD_r$. Therefore, there is a broad qualitative agreement between the predictions as far as fast decay is concerned; however, the identification of a slower mode of decay farther away from the boundary and the dependence of the corresponding length scale on the confinement radius has not been reported earlier. 

Our investigation of active particle-concave boundary interactions highlights the need for a further study of thermal-active noise coupling \cite{elgeti2015run, woillez2019activated, caprini2022role}. Ongoing research is extending these studies to convex and soft interfaces and various active particles. These efforts aim to deepen our understanding of waiting times at interfaces, which could be pivotal in designing active particle sorting in medical applications, steering particles for self-organization \cite{gompper20202020, araujo2023steering}, and advancing our knowledge of motility-induced phase transitions \cite{iyer2023dynamics}.


\begin{acknowledgments}
OV and BRS acknowledge Sushabhan Sadhukhan for support in bacterial culture. BRS acknowledges financial support from IIT Palakkad ERG Grant (2023-161-PHY-BRS-ERG-SP). AS and VVV acknowledges financial support from the NSM grant ($\mathrm{DST/NSM/R\&D\_HPC\_Applications/2021/29}$), IIT Palakkad ERG grant ($\mathrm{IITPKD/ICSR/2023/0.No.22}$), computational resources from Chandra and Param Vidya clusters at IIT Palakkad, and insightful discussions with Akshay Bhatnagar. We also gratefully acknowledge Elsa Baby for carefully proofreading the manuscript.

\end{acknowledgments}

\appendix
\section*{Appendix A: Experimental Details}
\subsubsection{E.coli Confinement in GUVs}The encapsulation of \emph{ Escherichia coli} bacteria in GUV was performed using the modified electroformation method \cite{angelova1986liposome,kuribayashi2006electroformation, takatori2020active}. \textit{Escherichia coli}(MTCC 739) was obtained from CSIR - Institute of Microbial Technology, Chandigarh culture collection and gene bank. 2mg/ml of electrically neutral 1,2-Dioleoyl-sn-glycero-3-phosphocholine (DOPC) is made in chloroform. 10$\mu$L of DOPC is evenly and uniformly coated on the conducting side of cleaned Indium Tin Oxide (ITO) coated plates. The plates are then kept in a desiccator for more than 2 hours to remove chloroform particles. After that a silicon spacer with thickness of about 2mm is sandwiched between ITO coated plates and 200 mM sucrose was added to the chamber. Then it is kept for electroformation with a square wave of 1Vpp at 10Hz in 30$^{\circ}$C for 45min. After 45 min of electroformation, a small amount of \textit{ E. coli} cultured over night in Luria-Bertani broth (LB) medium is introduced into the chamber. The setup is kept aside in the absence of voltage for 10 min. Then a square wave of 1Vpp at 10Hz was given for about 20 min. 200 mM glucose was used as the outside medium. The vesicles and bacteria sample after electroformation is observed with an Olympus IX83 Inverted Microscope in phase-contrast mode with 40X magnification.
\subsubsection{Image Analysis} Time lapse images of the motion of  \textit{ E. coli} inside the vesicle are recorded with 10 frames per second for 300s. The recorded images are analyzed using ImageJ software and bacteria are tracked using the TrackMate \cite{ershov2022trackmate} plugin available in ImageJ software.
\par 
The position coordinates of the bacterium are obtained from the TrackMate plugin, and the vesicle boundary coordinates are obtained in the ImageJ software. Then the shortest distance of the particle from the boundary is calculated for each radius of the vesicle and the positional probability distribution is plotted using the MATLAB software. 
\newline \textit{Estimation of $D_r$ and $D_t$}: 
The rotational diffusion coefficient of bacteria is estimated from the MSD of free bacteria prepared under the same conditions as confined (see Fig.S3, SM). The effective translational diffusion coefficient $D_{\rm eff}$ is obtained by analyzing the diffusive regime of the MSD on a long time scale and $D_r$ is found from it using the formula $D_r = u_0^2/2D_{\rm eff}$\cite{celani2010bacterial} in $d=2$(assuming that the thermal translational diffusion coefficient $D_t\ll D_{\rm eff}$). The propulsion speed of the bacteria used in our experiments was measured to be $u_0=12.25\pm 6.32 \mu$m$s^{-1}$. This analysis gives the value $D_r\simeq 0.37$s$^{-1}$ for the confined bacteria, from which the effective hydrodynamic radius $a=(k_BT/8\pi\eta_s D_r)^{1/3}$ for the bacterium is found (here, $\eta_s$ is the coefficient of viscosity of the sucrose solution \cite{swindells1958viscosities}) and has the value $a\simeq 0.71\mu$m. 

The thermal translational diffusion coefficient $D_t$ for the bacterium is now estimated using the standard Stokes-Einstein relation for a spherical colloidal particle. For radius $a=0.71\mu$m and $T=298 K$, we find $D_t=k_BT/6\pi\eta_s a\simeq 0.25 \mu$m$^2s^{-1}$.

\appendix
\section*{Appendix B: Mathematical Details}

\subsubsection{Derivation of the SFPE}

The joint probability density for the position and orientation variables together is $\Phi(r,\phi;\theta;t)$, which satisfies the FPE

\begin{equation}
    \frac{\partial \Phi(r,\phi;\theta;t)}{\partial t} = - \boldsymbol{\nabla} \cdot \mathbf{J}_t + D_r\frac{\partial^2 \Phi}{\partial\theta^2}
    \label{eq10}
\end{equation}
where 
\begin{equation}
    \mathbf{J}_t = -D_t \boldsymbol{\nabla} \Phi + \mathbf{u} \Phi
    \label{eq11}
\end{equation}
is the probability current density vector for translational motion, where the active velocity ${\bf u}=u_0(\cos\theta,\sin\theta)$. Now, expand $\nabla\cdot {\bf J}_t=P(\nabla\cdot {\bf u})+{\bf u}\cdot \nabla \Phi-D_t\nabla^2 \Phi$ and express each term in plane polar coordinates: 

\begin{eqnarray}
{\nabla}\cdot {\bf u}=\frac{1}{r}\frac{\partial}{\partial r}(ru_r)+\frac{1}{r}\frac{\partial u_{\phi}}{\partial \phi}\nonumber\\
{\bf u}\cdot\nabla P=u_r\frac{\partial P}{\partial r}+\frac{u_{\phi}}{r}\frac{\partial P}{\partial \phi}
\label{eq12}
\end{eqnarray}
where 
\begin{equation}
u_r=u_0 \cos\chi~~~~; u_\phi=u_0\sin\chi~~~(\chi=\theta-\phi), 
\label{eq13}
\end{equation}
and the two-dimensional Laplacian $\nabla^2=\partial^2/\partial r^2+r^{-1}\partial/\partial r+r^{-2}\partial^2/\partial \phi^2$. Considering the invariance of the boundary condition with respect to in-plane rotations, we expect the stationary state distribution to depend on $\theta$ only through the combination $\theta-\phi$, i.e., $\Phi(r,\phi;\theta)\to \Pi(r,\phi;\theta-\phi)$. Consequently,  the above derivatives are transformed as $\partial/\partial \phi\to \partial/\partial \phi-\partial/\partial \chi$ and $\partial/\partial \theta\to \partial/\partial \chi$. From the radial symmetry of the problem, the stationary distribution cannot explicitly depend on $\phi$ either, hence all $\phi$ derivatives identically vanish and $\Pi(r,\phi;\theta-\phi)\to \Pi(r,\chi)$. 

After substituting Eq.~\ref{eq11}, Eq.~\ref{eq12} and Eq.~\ref{eq13} in Eq.~\ref{eq10}, putting the time derivative to zero in stationary state and implementing the replacement $\Phi\to\Pi$, we arrive at the stationary FPE\cite{malakar2018steady}

\begin{eqnarray}
D_t\nabla_r^2 \Pi+\left(D_r+\frac{D_t}{r^2}\right)\frac{\partial^2 \Pi}{\partial \chi^2}+\frac{u_0}{r}\sin\chi\frac{\partial \Pi}{\partial \chi}-\nonumber\\ u_0\cos\chi\frac{\partial \Pi}{\partial r}=0
    \label{eq14}
\end{eqnarray}
where $\nabla^2_r=\partial^2/\partial r^2+r^{-1}\partial/\partial r$ is the radial part of the two-dimensional Laplacian operator. 
The above SFPE can be conveniently expressed in dimensionless form by defining the dimensionless radial coordinate $\xi=r/R$ and the two dimensionless ratios defined in Eq.~\ref{eq5}. The corresponding distribution $f(\xi,\chi)=R^2\Pi(r,\chi)$ satisfies the equation

\begin{equation}
\eta\frac{\partial^2 f}{\partial \xi^2}+\left(1+\frac{\eta}{\xi^2}\right)\frac{\partial^2 f}{\partial\chi^2}+\left(\frac{\eta}{\xi}-{\tilde u}\cos\chi\right)\frac{\partial f}{\partial \xi}+\frac{{\tilde u}}{\xi}\sin\chi\frac{\partial f}{\partial \chi}=0
\label{eq15}
\end{equation}
which is normalized as $\int_{-\pi}^{\pi}d\chi\int_{0}^{ }d\xi\xi f(\xi,\chi)=1$. The radial part of the probability current density is 

\begin{equation}
J_r=-D_t\frac{\partial P}{\partial r}+u_rP, 
\label{eq16}
\end{equation}
which, after non-dimensionalisation, and using Eq.~\ref{eq13}, becomes 

\begin{equation}
j(\xi,\chi)={\tilde u}\cos\chi f-\eta \frac{\partial f}{\partial \xi}
\label{eq17}
\end{equation}

in terms of which the dimensionless SFPE in Eq.~\ref{eq15} can be rewritten as 
\begin{equation}
\left(\xi+\frac{\eta}{\xi}\right)\frac{\partial^2 f}{\partial \chi^2}-\frac{\partial}{\partial \xi}(\xi j)+{\tilde u}\frac{\partial}{\partial \chi}(\sin\chi f)\simeq 0
\label{eq18}
\end{equation}
Next, we put $\xi=1+z$, where $z$ is the radial distance from the circular boundary. Substituting this expression in Eq.~\ref{eq15} and after keeping only the leading terms in the resulting expansions, we arrive at the approximate linearized SFPE in Eq.~\ref{eq4}.

\subsubsection{On the roots of the equation $\Gamma(s)=0$}
A graphical analysis (Fig.~\ref{figsimtheoryApp} a) suggests that the function $\Gamma(s)$ (Eq.~\ref{eq30}) two negative zeroes and one positive zero, all on the real axis, which we denote by $s_i (i=0,1,2)$: we adopt the convention that $s_0$ is the larger of the negative roots, $s_1$ is the other negative root and $s_2$ is the positive root. In general, the roots can be determined numerically, but in the limit $\eta\to 0$, approximate asymptotic expressions can be obtained for the same.

When $\eta\to 0$, from Eq.~\ref{eq30}, the equation $\Gamma(s)=0$ can be immediately simplified to 

\begin{equation}
\eta^2(s^3-s^2)-{\tilde u}^2{\overline\alpha}_1^2 s-{\tilde u}^2(1-{2\ \overline\alpha}_1^2)=0.
\label{eq31}
\end{equation}
Let us first neglect the first term in comparison with the other two, which leads to the solution $s\simeq s_0$, where $s_0=2-{\overline\alpha}_1^{-2}$. Note that, since $|{\overline\alpha}_1|<1$,  $s_0$ is negative, and independent of $\eta$. Two more real solutions exist for Eq.~\ref{eq31}, both of which are proportional to $\eta^{-1}$ in the limit $\eta\to 0$. To show this, we define ${\tilde s}=\eta s$ and substitute in Eq.~\ref{eq31}, which leads to the equation
\begin{equation}
\frac{1}{\eta}\{{\tilde s}^3-{\tilde u}^2{\overline\alpha}_1^2{\tilde s}\}-\{{\tilde u}^2(1-{2\ \overline\alpha}_1^2)+{\tilde s}^2\}=0
\label{eq32}
\end{equation}
If ${\tilde s}$ is considered independent of $\eta$, then the first term dominates over the second in the limit $\eta\to 0$. Ignoring the second term in Eq.~\ref{eq32}, we thus identify two more roots, i.e.,  ${\tilde s}\simeq \pm {\tilde u}{\overline\alpha}_1$ (apart from ${\tilde s}=0$, which is essentially the same as the root identified earlier, i.e.,  $s_0$,  which is independent of $\eta$). Therefore in this limit, the three solutions of Eq.~\ref{eq31} are given by the following expressions. 

\begin{equation}
s_0\sim 2-\frac{1}{\overline{\alpha}_1^2}; ~~s_{1}\sim -s_*; ~~~~s_2\sim s_*~~~ \bigg(s_*\equiv \frac{{\tilde u}{\overline{\alpha}}_1}{\eta}\bigg). 
\label{eq33}
\end{equation}
Fig.~\ref{figsimtheoryApp}~(b) and (c) show comparisons between the zeroes of $\Gamma(s)$ as determined numerically using a Newton-Raphson algorithm, versus the approximate analytical estimates given in Eq.~\ref{eq33} for a range of values of $\eta$. The agreement is found to be very good. 
\begin{figure}[htp!]
	\centering  \includegraphics[width=1\columnwidth]{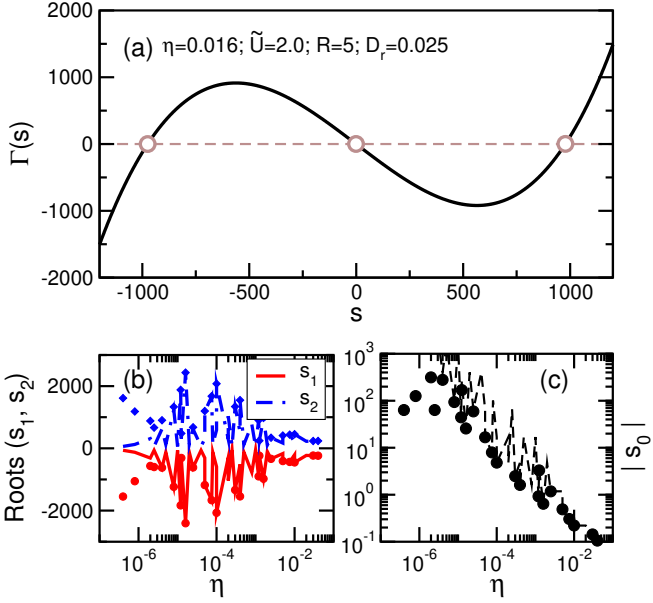}   
	\caption{(a) The function $\Gamma(s)$ (Eq.~\ref{eq30} is plotted against $s$ for $R=5$, $D_r=0.025$, which corresponds to ${\tilde u}=2.0$ and $\eta=0.016$. (b) and (c) show the roots of the equation $\Gamma(s)=0$ obtained using Newton-Raphson's method (solid line) and the approximate analytical solution provided in Eq.\ref{eq33} (symbols), as a function of $\eta$ and fixed ${\tilde u}=2.0$. Here, $s_1<0$ and $s_2>0$.}
	\label{figsimtheoryApp}
\end{figure}

\subsubsection{Inverse Laplace transform of Eq.~\ref{eq7}}

In terms of the three roots of the equation $\Gamma(s)=0$, the inverse Laplace transform of the expression in Eq.~\ref{eq7} can be expressed as 

\begin{equation}
P(z)=\sum_{i=0}^{2}{\rm Res}({\tilde P}, s_i)e^{s_i z}
\label{eq34}
\end{equation}
where ${\rm Res}({\tilde P},s_i)$ denote the residues of ${\tilde P}(s)$ at $s=s_i$ for $i=0,1,2$. The explicit expressions for the residues are 
\begin{eqnarray}
{\rm Res}({\tilde P},s_0)=\frac{A(s_0)P(0)-{\tilde u}(s_0-1)\lambda_1(0)}{\eta(s_0-s_1)(s_0-s_2)}\nonumber\\
{\rm Res}({\tilde P},s_1)=\frac{A(s_1)P(0)-{\tilde u}(s_1-1)\lambda_1(0)}{\eta(s_1-s_0)(s_1-s_2)}\nonumber\\
{\rm Res}({\tilde P},s_2)=\frac{A(s_2)P(0)-{\tilde u}(s_2-1)\lambda_1(0)}{\eta(s_2-s_0)(s_2-s_1)}\nonumber\\
\label{eq35}
\end{eqnarray}
The $s_2$ term is exponentially increasing with $y$, which is not in agreement with experimental data. Therefore, we equate the corresponding residue to zero, which leads to the following relation between the boundary values $P(0)$ and $\lambda_1(0)$:
\begin{equation}
{\tilde u}{\lambda}_1(0)=P(0)\frac{A(s_2)}{(s_2-1)}
\label{eq36}
\end{equation}
In the limit $\eta\to 0$, $s_2\sim s^*$, and $s^{*}\gg 1$(Eq.~\ref{eq33})and $A(s_2)\sim \eta s_2^2$. Hence, the above relation simplifies to $\lambda_1(0)\simeq {\overline\alpha}_1 P(0)$, which is consistent with the mean-field assumption already made (Eq.~\ref{eq25++}, also Eq.~\ref{eq27+}). 

After using the relation Eq.~\ref{eq36} in Eq.~\ref{eq34} and Eq.~\ref{eq35}, we arrive at the bi-exponential expression for $P(z)$ in Eq.~\ref{eq9}. The constants $C$ and $D$ are determined by the residues in Eq.~\ref{eq35}. 


\subsubsection{The reflecting boundary condition and the fast exponential decay of $P(z)$}
 The smaller length scale $\zeta_1$ has an intuitive physical interpretation. To see this, let us integrate Eq.~\ref{eq4} over $\chi$. The first term vanishes on account of periodicity with respect to $\chi$, and the resulting equation is 

\begin{equation}
\frac{d}{dz}({\overline j}-\eta P)=0
\label{eq9+}
\end{equation}
where $\overline{j}(z)=\eta dP/dz+{\tilde u}\langle \cos\chi\rangle_z P(z)$ is the angular average of the radial probability current density in Eq.~\ref{eq4+}, which satisfies the boundary condition $\overline{j}(0)=0$. Let us now assume that $\eta\ll \langle \cos \chi\rangle_z$, in which case Eq.~\ref{eq9+} simplifies to ${\overline j}(z)=0$, after taking into account the boundary condition. A mean-field approximation suggests the replacement $\langle \cos\chi\rangle_z\to {\overline \alpha}_1$ (Eq.~\ref{eq27}), which, when implemented, leads to the  exponential solution $P(z)\propto \exp(-{\tilde u}{\overline \alpha}_1 z/\eta)$, matching the second term in Eq.~\ref{eq9} (which is dominant at small $z$). This is consistent, because simulation results show that, in the small $z$-regime,  the conditional mean of $\cos\chi$ can be approximated well by a constant value (Fig.~\ref{figsimtheory} b).

\newpage   


\bibliography{reference}
\newpage
\section*{Supplementary Information (SI)}

\subsection*{1. Supplementary Figures}
The trajectory of the dead bacterium, the $\zeta_1(R)$ in log-log plot showing the power law behaviour and average MSD of the free bacteria is shown in the Fig. \ref{sup1}, Fig. \ref{sup2} and Fig. \ref{sup3} respectively.

\begin{figure}[ht!]
   \centering  \includegraphics[width=0.95\columnwidth]{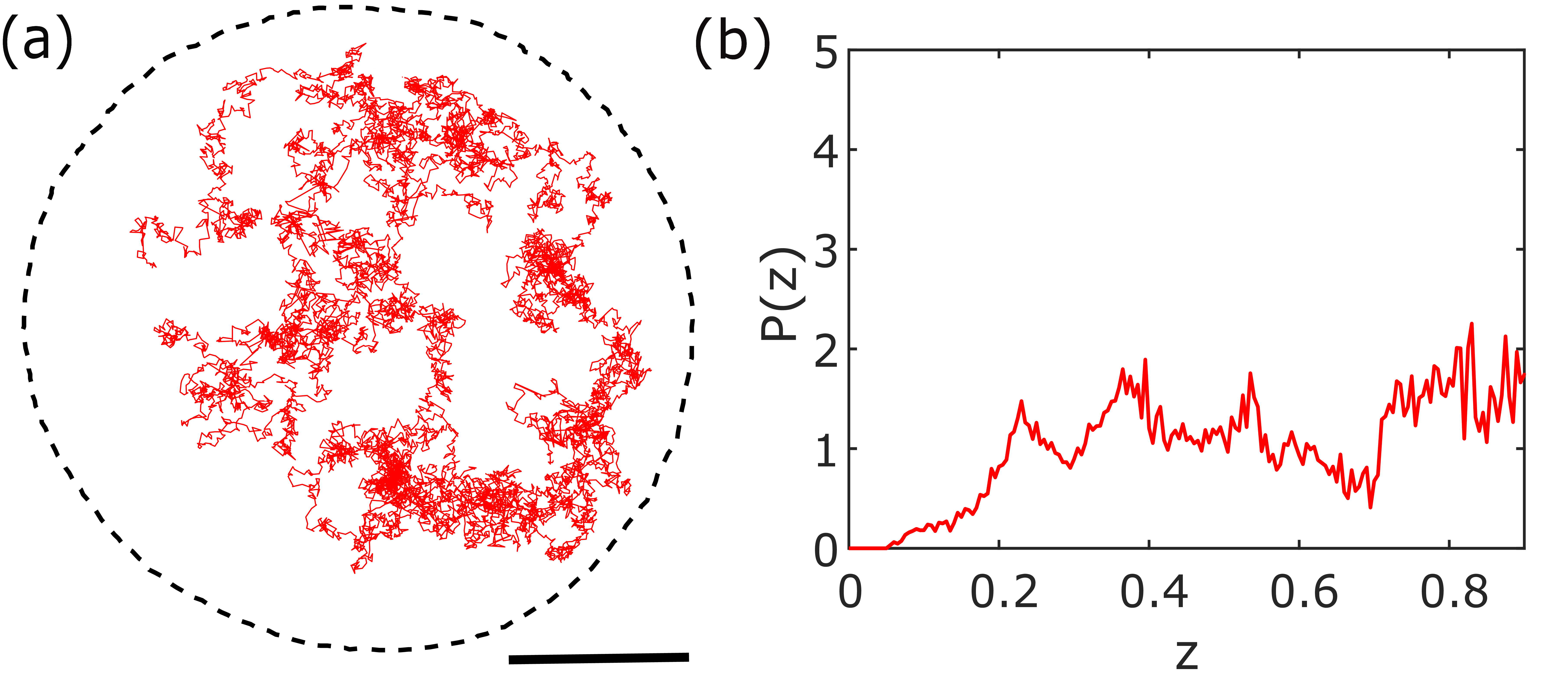}

    \caption{a) Trajectory(red line) of the dead bacterium inside a vesicle of radius 18$\mu$m. Vesicle boundary is given in black dashed line. The scalebar is 10$\mu$m. b) The probability distribution of the dead bacterium inside vesicle of radius 18$\mu$m.The trajectory of a single dead bacterium inside a vesicle of radius 18$\mu$m is found to be a uniform distribution. }
    \label{sup1}
\end{figure}

\begin{figure}[ht!]
   \centering
	\includegraphics[width=0.95\columnwidth]{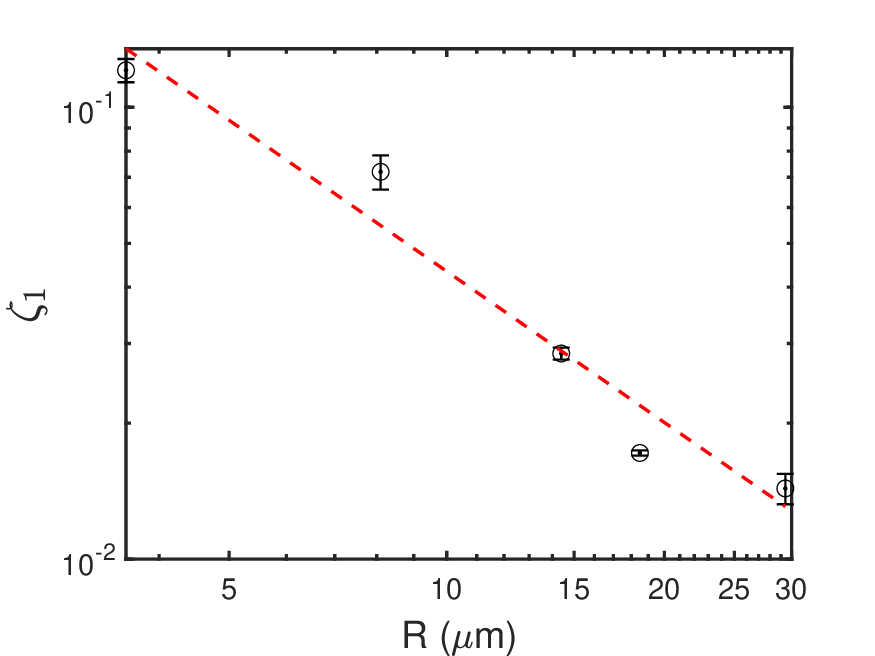}

    \caption{Plot showing the dependence of $\zeta_1$ on R in log-log scale. The red dashed line is the linear fit with slope -1.11}
    \label{sup2}
\end{figure}

\begin{figure}
   \centering
	\includegraphics[width=0.95\columnwidth]{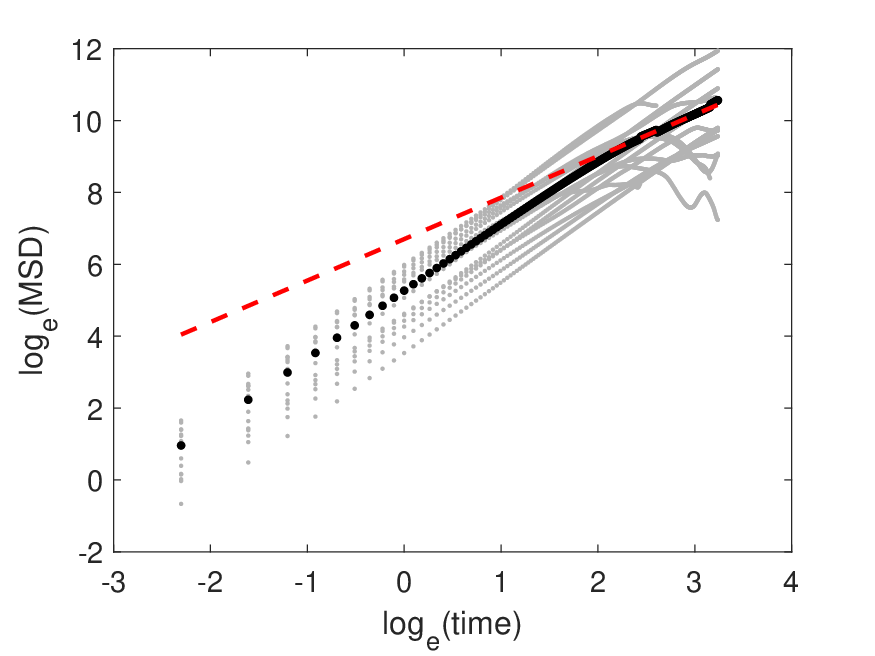}

    \caption{Averaged MSD plot of free bacteria in sucrose solution(Black dots). The red dashed line is a straight line fit with slope $\approx$ 1 to the diffusive regime on a longer timescale. Grey dots corresponds to the MSD of individual bacterium.   }
    \label{sup3}
\end{figure}

\subsection*{2. Supplementary Movies}
The motion of Bacterium confined within a Giant Unilamellar Vesicle(GUV) is shown in the following links.
\href{https://youtu.be/XC3cMcrKzbA?si=W0TIFGGHu9N4jaGJ}
 {Supplementary Movie-1}, \href{https://youtu.be/HO8hVrsazIk?si=QX7nbkYkZClGQgO1}{Supplementary Movie-2}\\
Supplementary Movie 1 : Time-lapse images of a bacterium encapsulated inside vesicles of different radius. Top Panel : Phase contrast images, Bottom Panel shows the corresponding tracks. 

Supplementary Movie 2 : Time-lapse images of a dead bacterium encapsulated inside vesicles.

 \subsection*{3. Supplementary Table}
 The table (TABLE-1) provides the comparison of length scales between experiments and theoretical predictions.

    \renewcommand{\arraystretch}{2.0}
\begin{table*}[t]
    \centering
    \begin{tabular}{|c|c|c|c|c|c|c|c|c|}
        \hline
        \multirow{2}{*}{$R$($\mu$m)} & \multirow{2}{*}{$\tilde{u}$} & \multirow{2}{*}{$\eta$} & \multicolumn{3}{|c|}{$\zeta_1$} & \multicolumn{3}{|c|}{$\zeta_2$} \\
        \cline{4-9}
        & & & Exp. & SFPE1 & SFPE2 & Exp. & SFPE1 & SFPE2 \\
        \hline
        3.6 & 9.197 & 0.05214 &  0.1644& 0.00420 & 0.0058 & - & 1.12 & 6.83 \\
            \hline                                                                                          
        8.1 & 4.087 & 0.01030 & 0.0727 & 0.00218 & 0.0030 & 0.3965 & 1.24 & 3.30 \\
            \hline                                                                                              
        18.5 &1.790 & 0.00197 & 0.0188 & 0.00093 & 0.0013 & 0.5577 & 1.54 & 1.20  \\
            \hline                                                                                          
        29.4 &1.130 & 0.00078 & 0.0128 & 0.00048 & 0.0008 & -      & 1.76 & 0.50\\
        \hline
    \end{tabular}

\vspace{0.2cm}
   \caption{{\it Comparison of the length scales $\zeta_1$ and $\zeta_2$ obtained from experiments with theoretical predictions from the ABP model, based on the SFPE. Here Exp. corresponds to experimental data. The SFPE predictions are obtained from the numerical roots of the equation $\Gamma(s)=0$ (Eq.~22) using (a) the boundary value $\alpha_1(0)$ obtained from simulations in place of ${\overline{\alpha}}_1$ (SFPE1) and (b) the semi-mean-field prediction in Eq.~20 (SFPE2). Appropriate combinations of $\tilde{u}$ and $\eta$ were chosen to compare to match theoretical predictions with experimental data.}}
\end{table*}
\newpage
\subsection*{Reflecting Boundary Condition}

In the simulations, we keep track of the position of the particle. At a given time step, if the radial distance of the ABP is greater than the radius of the confinement, we apply the reflecting boundary condition (see below equation and Fig. \ref{sup3}) after finding the point of contact at the boundary (xB,yB) - see Figure. In the equation, the subscript “ref” corresponds to the reflected velocity or position and “old” corresponds to the velocity at the crossing timestep. 
\begin{eqnarray*}
\vec{v}_{ref} &=& \vec{v}_{old} - 2\left ( \vec{v}_{old} \cdot \hat{r}_B \right )\hat{r}_B\\
\vec{r}_{ref} &=& \vec{r}_2 + \vec{v}_{ref} dt
\end{eqnarray*}

\begin{figure}[ht!]
   \centering
	\includegraphics[width=0.5\columnwidth]{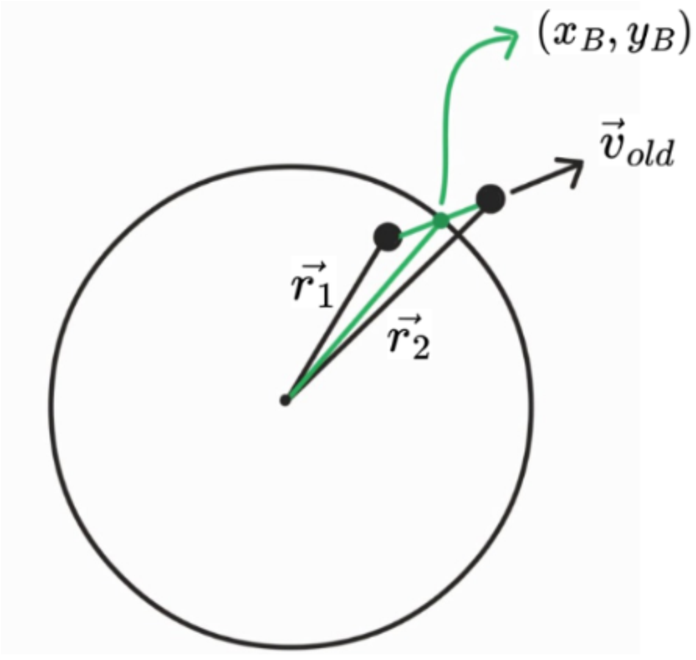}    
    \caption{Reflecting Boundary Condition (RBC) applied for particles moving out of the confining circular boundary.}
    \label{sup2}
\end{figure}

\end{document}